\documentclass{article}
\usepackage{dcase2020,amsmath,graphicx,url,times,booktabs, tabularx}

\usepackage{amssymb}
\usepackage{multirow}

\usepackage{subfigure} 
\graphicspath{{./image/}}

\title{On Multitask Loss Function for Audio Event Detection and Localization}

\name{Huy Phan$^{*1}$,
	Lam Pham$^{2}$,
	Philipp Koch$^{3}$,
	Ngoc Q. K. Duong$^{4}$,
	Ian McLoughlin$^{5}$,
	Alfred Mertins$^{3}$
}
\address{$^1$ School of Electric Engineering and Computer Science, Queen Mary University of London, UK \\                 
	$^2$ School of Computing, University of Kent, UK \\
	$^3$ Institute for Signal Processing, University of L\"ubeck, Germany \\
	$^4$ InterDigital R\&D France, France \\
	$^5$ Singapore Institute of Technology, Singapore \\
	$^*$Corresponding email: {\texttt{h.phan@qmul.ac.uk}}
}

\begin{document}

\maketitle

\begin{sloppy}

\begin{abstract}
Audio event localization and detection (SELD) have been commonly tackled using multitask models. Such a model usually consists of a multi-label event classification branch with sigmoid cross-entropy loss for event activity detection and a regression branch with mean squared error loss for direction-of-arrival estimation. In this work, we propose a multitask regression model, in which both (multi-label) event detection and localization are formulated as regression problems and use the mean squared error loss homogeneously for model training. We show that the common combination of heterogeneous loss functions causes the network to underfit the data whereas the homogeneous mean squared error loss leads to better convergence and performance. Experiments on the development and validation sets of the DCASE 2020 SELD task demonstrate that the proposed system also outperforms the DCASE 2020 SELD baseline across all the detection and localization metrics, reducing the overall SELD error (the combined metric) by approximately $10\%$ absolute.
\end{abstract}


\vspace{-0.15cm}
\section{Introduction}
\label{intro}
\vspace{-0.15cm}
Extended from active research on sound (audio) event detection, sound event localization and detection (SELD) task \cite{Politis2020, Adavanne2018} entangles the \emph{what} and \emph{where} questions about occurring sound events. That is, it aims to determine the identities of the events and their spatial locations/trajectories simultaneously. Solving the SELD task would enable a wide range of novel applications in surveillance, human-machine interaction, bioacoustics, and healthcare monitoring, to mention a few.

The joint SELD task can be divided and conquered individually by two separate models, one for sound event detection (SED) \cite{Phan2019, Cakir2017, McLoughlin2017} and the other for sound source localization (SSL) \cite{Ma2018, Chakraborty2014}. The two-stage approach presented in \cite{Cao2019} can be also considered to belong to this line of work. Dealing with the joint task in a single model has been known to be more challenging. Three main approaches have been proposed, including sound-type masked SSL \cite{Ma2018}, spatially masked SED \cite{Trowitzsch2020}, and joint SELD modeling \cite{He2018,Adavanne2018}. Joint sound event detection and localization modeling with multitask deep learning has been most commonly adopted in the latest DCASE challenge \cite{Grondin2019,Cordourier2019,Kapka2019, Adavanne2018}, demonstrating encouraging results.

In the joint modeling approach with a multitask model, the sigmoid cross-entropy (CE) loss is typically used for event detection (via classification) to handle possible multi-label due to occurrences of multiple events while the mean squared error (MSE) loss is often employed for direction-of-arrival (DOA) estimation (via regression). These two losses are usually associated with different weights and then combined to make the total loss for network training. However, there exist no established rules to set the weights for the losses; more often than not, they are set with some trivial weights without a clear justification. For example, while the DCASE 2019 baseline weighted the MSE loss 50 times larger than that of the sigmoid CE loss, the current DCASE 2020 baseline even enlarges this multiplication to 1000 times. Furthermore, the two different types of loss functions might progress at different rates and might not converge synchronously, making the fixed weights suboptimal. We will empirically show in a controlled experiment that, for this joint modeling task, the classification based on the CE loss usually experiences \emph{underfitting} when being optimized jointly with regression based on the MSE loss.

In order to avoid such an issue, we alternatively propose to formulate both the SED and SSL subtasks as regression problems and homogeneously use the MSE loss for both of them. The proposed multitask-regression network features a recurrent convolutional neural network (CRNN) architecture coupled with self-attention mechanism \cite{Vaswani2017}. Experiments on the development set of the DCASE 2020 Task 3 show that the proposed multitask-regression network results in better generalization than the networks using the combination of the CE loss and the MSE loss. Furthermore, evaluation on the development and evaluation data of the challenge shows that the proposed network outperforms the DCASE 2020 SELD baseline across all the evaluation metrics, some with a large margin. 

\vspace{-0.25cm}
\section{The proposed network}
\vspace{-0.15cm}
The proposed network is illustrated in Figure \ref{fig:network}. The network receives time-frequency input $\mathbf{S} \in \mathbb{R}^{T\times F \times C}$ of $T$ frames, $F$ frequency bins, and $C$ channels. The convolutional part of the network consists of six convolutional layers each of which is followed by a max pooling layer except the first one. We assume that the early convolutional layers are crucial for feature learning, the network is designed to have the first two convolutional layers back-to-back. In order, the six convolutional layers accommodate $\{64, 64, 128, 128, 256, 256\}$ filters, respectively, with a common kernel size of $3\times3$ and the stride of $1\times1$. The gradually increasing numbers of filters in the later convolutional layers are to compensate for their smaller feature maps in the frequency dimension. Zero-padding (i.e. \emph{SAME} padding) is used in order to preserve the temporal size. After convolution, batch normalization \cite{Ioffe2015} is applied on the feature maps, followed by Rectified Linear Units (ReLU) activation \cite{Nair2010}.

The max pooling layers, except the first one, have a common kernel size of $1 \times 2$ to reduce the input size by half in the frequency dimension and, by doing so, gain frequency equivariance in the induced feature maps while keeping the temporal size unchanged. Particularly, the pooling kernel size of the first max pooling layer (\emph{max pool 2}, cf. Figure \ref{fig:network}) is set to $5 \times 2$ in order to reduce the time dimension to $\frac{T}{5}$ to match the frame resolution (100 ms) for computing the evaluation metrics. 

Passing through the convolutional block, the input is transformed into a feature map of size $\frac{T}{5}\times 2 \times 256$ which is reshaped to form a sequence of feature vectors $(\mathbf{x}_1, \mathbf{x}_2, \ldots, \mathbf{x}_{\frac{T}{5}})$ where $\mathbf{x}_i \in \mathbb{R}^{512}$, $1 \le i \le \frac{T}{5}$. A bidirectional recurrent neural network (biRNN) is then employed to iterate through the sequence and encode it into a sequence of output vectors $(\mathbf{z}_1, \mathbf{z}_2, \ldots, \mathbf{z}_{\frac{T}{5}})$. The biRNN is realized by Gated Recurrent Unit (GRU) cells with the hidden size of 256.
To further improve encoding the context around a feature $\mathbf{z}_i$, self-attention mechanism \cite{Vaswani2017} is used. The vectors $(\mathbf{z}_1, \mathbf{z}_2, \ldots, \mathbf{z}_{\frac{T}{5}})$ can be viewed as as a set of \emph{key-value} pairs $(\mathbf{K},\mathbf{V})$. In the context of this work, both the keys and values coincide to $\mathbf{Z}$ (the concatetation of the $\mathbf{z}_1, \mathbf{z}_2, \ldots, \mathbf{z}_{\frac{T}{5}}$ vectors). We adopt the scaled dot-product attention as in \cite{Vaswani2017}, i.e. the attention output at a time index is a weighted sum of $\mathbf{z}_1, \mathbf{z}_2, \ldots, \mathbf{z}_{\frac{T}{5}}$ where the weights are determined as
\begin{align}
Attention(\mathbf{Q}, \mathbf{K}, \mathbf{V}) = softmax(\frac{\mathbf{Q}\mathbf{K}^{\mathsf{T}}}{\sqrt{d_k}})\mathbf{V}.
\end{align}
Here, $\mathbf{Q}$ is the \emph{query} \cite{Vaswani2017} and also coincides to $\mathbf{Z}$ in the context of this work, i.e. $\mathbf{Q} \equiv \mathbf{K} \equiv \mathbf{V} \equiv \mathbf{Z}$. $d_k$ is the extra dimension into which $\mathbf{Q}$, $\mathbf{K}$ are transformed before the dot product to prevent the inner product from becoming too large. $d_k$ is set to 64 in this work.
\begin{figure} [!t]
	\centering
	\includegraphics[width=0.9\linewidth]{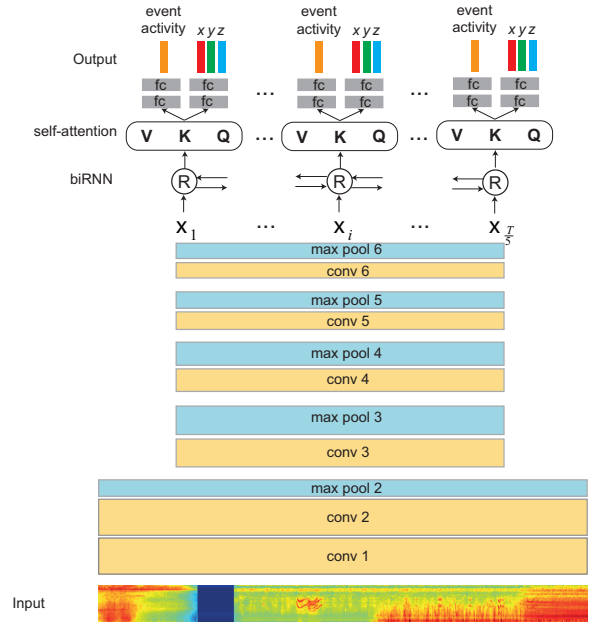}
	\vspace{-0.25cm}
	\caption{Overview of the proposed multitask regression self-attention CRNN.}
	\label{fig:network}
\vspace{-0.25cm}
\end{figure}

At each time index, the SED and SSL subtasks are accomplished via two network branches, each consisting of two fully connected (fc) layers with 512 units each. The first branch's output layer has $\mathcal{Y}$ units with \emph{sigmoid} activation to perform event activity classification/regression of $\mathcal{Y}$ classes. The second branch has $3\mathcal{Y}$ units with \emph{tanh} activation to regress for the target events' DOA trajectories. Normally, when the sigmoid CE loss is used for event activity classification and the MSE loss is used for the DOA estimation, the network is trained to minimize the following weighting loss:
{
	\small
\begin{align}
\mathcal{L}_{\text{CE}+\text{MSE}}(\Theta)\!=\!&-w_{\text{CE}}\!\sum_{n=1}^N\!\sum_{t=1}^{\frac{T}{5}} (\mathbf{y}_{nt}\!\log(\hat{\mathbf{y}}_{nt}\!)\!+\!(1\!-\!\mathbf{y}_{nt}\!)\log(1\!-\!\hat{\mathbf{y}}_{nt}\!)\!) \nonumber\\ & + w_{\text{MSE}}\sum_{n=1}^N\!\sum_{t=1}^{\frac{T}{5}}||\hat{\mathbf{d}}_{nt}(\Theta) - \mathbf{d}_{nt}||^2. \label{eq:ce_mse_loss}
\end{align}
}
Here, $\Theta$ denotes the network parameters and $N$ denotes the number of training examples. We use $\hat{\mathbf{y}}$ and $\mathbf{y}$ to denote the event activity output and grouthtruth, respectively. In addition, we used $\hat{\mathbf{d}} = (\hat{x},\hat{y},\hat{z})$ and $\mathbf{d} = (x,y,z)$ to denote the DOA estimation output and groudtruth in terms of Cartesian coordinates on the unit sphere, respectively. $w_{\text{CE}}$ and $w_{\text{MSE}}$ indicate the weights given to the corresponding losses.

On the other hand, when the MSE loss is used for both SED and SSL subtasks, the network is trained to minimize the total MSE loss of the two network branches without weighting:
{
\small
\begin{align}
\mathcal{L}_{\text{MSE}}(\Theta)\!=\!\sum_{n=1}^N\!\sum_{t=1}^{\frac{T}{5}} (||\hat{\mathbf{y}}_{nt}(\Theta)\!-\!\mathbf{y}_{nt}||^2 + ||\hat{\mathbf{d}}_{nt}(\Theta)\!-\!\mathbf{d}_{nt}||^2). \label{eq:mse_loss}
\end{align}
}


\vspace{-0.35cm}
\section{Experiments }
\vspace{-0.15cm}
\subsection{DCASE 2020 SELD dataset}
\vspace{-0.15cm}
The database used for the DCASE 2020 SELD task was synthesized in two spatial sound formats: (1) MIC - 4-channel microphone array extracted from a subset of 32-channel Eigenmike format and (2) FOA - 4-channel first-order Ambisonics extracted from a matrix of $4\times 32$ conversion filters. 714 sound examples from the published NIGENS General Sound Events Database \footnote{https://zenodo.org/record/2535878} of 14 event classes, including \emph{alarm, crying baby, crash, barking dog, running engine, burning fire, footsteps, knocking on door, female \& male speech, female \& male scream, ringing phone,} and \emph{piano}, were used for data creation. More information about the data synthesis can be found in \cite{Politis2020}. The database was split into eight sets, six of which were used as the development set and the remaining two were used as the evaluation set. 

\emph{Experiments on the development set:} We followed the challenge setup to conduct experiments on the development set. That is, the first set of the development data was used as the unseen data for testing purpose, the second set was used as the validation set for model selection, and the remaining four sets were used as the training data. 

\emph{Experiments on the evaluation set:} To assess performance on the evaluation set, two different systems were trained and submitted to the challenge. The first was trained using the first set of the development data as validation set for model selection and the remaining five sets as the training data ({\bf Submission 1}). The second was trained using the entire development data as the training data (i.e. without validation data for model selection) ({\bf Submission 2}).

\vspace{-0.25cm}
\subsection{Feature extraction}
\vspace{-0.15cm}
We extracted log-Mel magnitude spectrogram with a window size of 40 ms, 20 ms overlap, and 64 Mel-bands. To encode the phase information, for the FOA data, an acoustic intensity vector was extracted for each Mel-band, whereas, for the MIC data,  generalized-cross-correlation with phase-transform (GCC-PHAT) features were computed for each Mel-band. Overall, multi-channel images of size $3000\times64\times7$ and $3000\times64\times10$ were resulted for one-minute FOA and MIC recordings, respectively.
\setlength\tabcolsep{1.25pt}
\begin{table*}[t!]
	\caption{Results obtained by the proposed system and the DCASE 2020 baseline on the development and evaluation sets.}
	\vspace{-0.25cm}
	\footnotesize
	\begin{center}
		\begin{tabular}{>{\arraybackslash}m{0.9in}|>{\centering\arraybackslash}m{0.85in}|>{\centering\arraybackslash}m{0.85in}|>{\centering\arraybackslash}m{0.35in}>{\centering\arraybackslash}m{0.35in}>{\centering\arraybackslash}m{0.35in}>{\centering\arraybackslash}m{0.35in}>{\centering\arraybackslash}m{0.35in}|>{\centering\arraybackslash}m{0.35in}>{\centering\arraybackslash}m{0.35in}>{\centering\arraybackslash}m{0.35in}>{\centering\arraybackslash}m{0.35in}>{\centering\arraybackslash}m{0.35in}}
			\multicolumn{1}{c|}{} & \multirow{2}{*}{DOA loss (weight)} & \multirow{2}{*}{SED loss (weight)} & \multicolumn{5}{c|}{\bf FOA} & \multicolumn{5}{c}{\bf MIC}\\
			\cline{4-13}
			\multicolumn{1}{c|}{} &  & & $LE_{CD}$ & $LR_{CD}$ & $ER_{20^\circ}$ & $F_{20^\circ}$ & $SELD$ & $LE_{CD}$ & $LR_{CD}$ & $ER_{20^\circ}$ & $F_{20^\circ}$ & $SELD$ \\
			\hline
			\multicolumn{11}{l}{\bf Development results \hfill}\\
			\hline
			Val (DCASE2020) & MSE (1000) & CE (1) & $23.5^\circ$ & $62.0$ & $0.72$ & $37.7$ & $0.46$ & $27.0^\circ$ & $62.6$ & $0.74$ & $34.2$ & $0.48$ \\ 
			
			\emph{Val (CE+MSE)} & MSE (1000) & CE (1) &  $\it 16.1^\circ$ & $\it 51.7$ & $\it 0.83$ & $\it 41.4$ & $\it 0.50$ & $\it 16.5^\circ$ & $\it 51.1$ & $\it 0.82$ & $\it 42.6$ & $\it 0.49$ \\ 
			
			\emph{Val (CE+MSE)} & MSE (1) & CE (1) & $\it 24.1^\circ$ & $\it 67.6$ & $\it 0.78$ & $\it 42.8$ & $\it 0.45$ & $\it 27.9^\circ$ & $\it 66.6$ & $\it 0.86$ & $\it 34.7$ & $\it 0.50$ \\ 
			
			\bf Val (MSE) & MSE & MSE & $\mathbf{17.7^\circ}$ & $\mathbf{68.1}$ & $\mathbf{0.58}$ & $\mathbf{52.4}$ & $\mathbf{0.37}$ & $\mathbf{17.3^\circ}$ & $\mathbf{66.0}$ & $\mathbf{0.56}$ & $\mathbf{53.9}$ & $\mathbf{0.37}$ \\ 
			
			Test (DCASE2020) & MSE (1000) & CE (1) & $22.8^\circ$ & $60.7$ & $0.72$ & $37.4$ & $0.47$ &  $27.3^\circ$ & $59.0$ & $0.78$ & $31.4$ & $0.51$\\ 
			
			\emph{Test (CE+MSE)} & MSE (1000) & CE (1) & $\it 18.0^\circ$ & $\it 50.6$ & $\it 0.88$ & $\it 38.9$ & $\it 0.53$ &  $\it 16.7^\circ$ & $\it 53.6$ & $\it 0.81$ & $\it 44.3$ & $\it 0.48$\\ 
			\emph{Test (CE+MSE)} & MSE (1) & CE (1) & $\it 26.2^\circ$ & $\it 62.7$ & $\it 0.82$ & $\it 39.9$ & $\it 0.49$ &  $\it 28.3^\circ$ & $\it 60.0$ & $\it 0.93$ & $\it 31.2$ & $\it 0.54$\\ 
			\bf Test (MSE) & MSE & MSE & $\mathbf{19.0^\circ}$ & $\mathbf{65.6}$ & $\mathbf{0.60}$ & $\mathbf{49.2}$ & $\mathbf{0.39}$ &  $\mathbf{18.2^\circ}$ & $\mathbf{64.1}$ & $\mathbf{0.59}$ & $\mathbf{50.8}$ & $\mathbf{0.38}$\\ 
			\hline
			\multicolumn{11}{l}{\bf Evaluation results \hfill}\\
			\hline
			DCASE2020 & MSE (1000) & CE (1) & $20.5^\circ$ & $65.0$ & $0.66$ & $43.3$ & $0.42$ & $21.8^\circ$ & $65.9$ & $0.66$ & $44.0$ & $0.42$\\ 
			
			\bf Submission 1 & MSE & MSE & $\mathbf{16.8^\circ}$ & $\mathbf{69.8}$ & $\mathbf{0.52}$ & $\mathbf{57.8}$ & $\mathbf{0.33}$ & $\mathbf{14.6^\circ}$ & $\mathbf{68.2}$ & $\mathbf{0.55}$ & $\mathbf{58.8}$ & $\mathbf{0.34}$\\ 
			\bf Submission 2 & MSE  & MSE & $\mathbf{15.2^\circ}$ & $\mathbf{72.4}$ & $\mathbf{0.49}$ & $\mathbf{61.7}$ & $\mathbf{0.31}$ & $\mathbf{14.6^\circ}$ & $\mathbf{68.2}$ & $\mathbf{0.53}$ & $\mathbf{59.2}$ & $\mathbf{0.33}$\\ 
			\hline
		\end{tabular}
	\end{center}
	\label{tab:development_results}
	\vspace{-0.45cm}
\end{table*}

\vspace{-0.25cm}
\subsection{Parameters}
\vspace{-0.15cm}
Network implementation was based on \emph{Tensorflow} framework. We used spectrogram segments of size $T=600$ (equivalent to 12 seconds) as inputs. \emph{Dropout} rates of $0.5$, $0.1$, and $0.25$ were employed to regularize the convolutional layers, the biRNN, and the fully-connected layers, respectively.

The network was trained using \emph{Adam} optimizer \cite{Kingma2015} for 10000 epochs with a minibatch size of 64. Each spectrogram segment in a minibatch was randomly sampled from a 1-minute recording and augmented using spectrogram augmentation \cite{Park2019}. The learning rate was initially set to $2\times 10^{-4}$ and was exponentially reduced with a rate of $0.8$ after $200$, $600$, and $1000$ epochs. In addition, the first $10$ epochs were used as a warmup period in which the network was trained with a small learning rate of $2\times 10^{-5}$.

During training, the network snapshot that achieved the lowest combined SELD error rate on the validation set was retained for evaluation. The retained network was then evaluated on the test recordings with a 2-second segment at a time without overlap. To be able to analyze the effect of using different loss combinations in a controllable manner, no post-processing was carried out. Event activity was determined from the corresponding regression/classification output using a threshold of $0.5$. 

\vspace{-0.25cm}
\subsection{Evaluation metrics}
\vspace{-0.15cm}
The DCASE 2020 challenge evaluated the performance of the SED subtask using localization-aware detection error rate ($ER_{20^\circ}$) and F-score ($F_{20^\circ}$) with a threshold of $20^\circ$ in one-second non-overlapping segments. For sound event localization, errors only between same-class predictions and references were considered. The class-aware localization error ($LE_{CD}$) and its corresponding recall ($LR_{CD}$) were employed for evaluating localization outputs and were also computed in one-second non-overlapping segments. In addition, we also computed the combined SELD error metric: 
\vspace{-0.15cm}
{
	\small
\begin{align}
SELD\!=\!\frac{1}{4}(\!ER_{20^\circ} + (1\!-\!F_{20^\circ})+\frac{LE_{CD}}{180} + (1\!-\!LR_{CD})\!)
\end{align}
}
to give an overall picture about a system.

\vspace{-0.25cm}
\subsection{Experimental results}
\vspace{-0.15cm}
\subsubsection{Influence of the loss functions}
\vspace{-0.15cm}
It is a rule of thumb that the CE loss is preferred over the MSE loss for a classification task since it, in general, leads to quicker learning through gradient descent, at least theoretically \cite{Nielsen2019}. However, when it is used in combination with the MSE task as in (\ref{eq:ce_mse_loss}) as commonly used for joint SELD, it apparently underfits the data as evidenced in Figure \ref{fig:loss}. When an equal weight is used for the two losses in (\ref{eq:ce_mse_loss}), i.e. $w_{\text{CE}} = w_{\text{MSE}} = 1$, the CE loss (cf. Figure \ref{fig:loss} (c)) and the SED error (cf. \ref{fig:loss} (d)) are hard to be reduced on both the training and test data (note the scale of the CE loss in Figure \ref{fig:loss} (c) is much larger than that of the MSE loss in Figure \ref{fig:loss} (a)). The underfitting effect on the SED subtask is even worse under the skewed weighting scheme used in the DCASE 2020 baseline \cite{Politis2020}, i.e. the MSE loss was given a weight of 1000.0 and the CE loss was given a weight of 1.0, since in this case the network further prioritizes optimizing the MSE loss over the CE one. We speculate that a similar phenomenon happened to the DCASE 2020 baseline as it results in limited performance on the SED subtasks (cf. Table \ref{tab:development_results}).

In contrast, when the MSE loss is used for both the SED and SSL subtasks as in (\ref{eq:mse_loss}), the SED performance is improved significantly (cf. Figure \ref{fig:loss} (d)) while the DOA estimation performance remains comparable to the case of MSE+CE combination (cf. Figure \ref{fig:loss} (b)). These results suggest that the SELD multitask network learns easier when a homogeneous loss is used for all the subtasks than when heterogeneous losses are combined. Although we cannot conclude that the MSE loss is the optimal loss for SELD multitask modeling, these results urge the quest for one in future work. 
\begin{figure} [!t]
	\centering
	\includegraphics[width=0.95\linewidth]{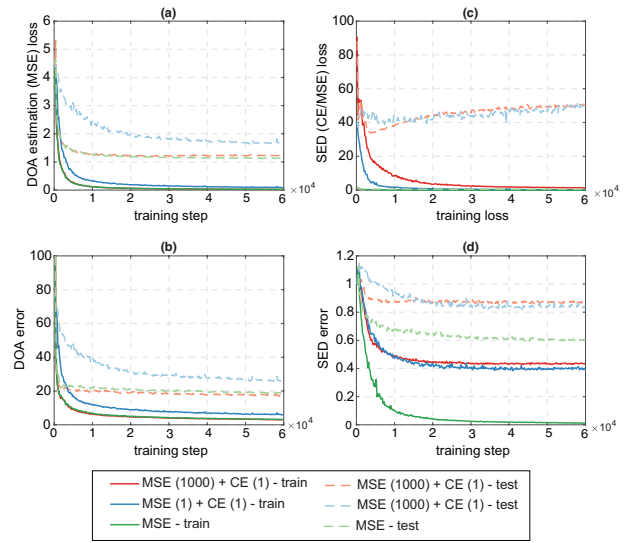}
	\vspace{-0.25cm}
	\caption{Variation of the MSE loss, the CE loss, the DOA error, and the SED error on the training and test sets of the DCASE 2020 development data with different loss combinations. (a) The MSE loss, (b) the CE loss, (c) the DOA error, and (d) the SED error. The number in bracket indicates the weight assigned to the corresponding loss.}
	\label{fig:loss}
	\vspace{-0.25cm}
\end{figure}

\vspace{-0.25cm}
\subsubsection{SELD performance}
\vspace{-0.15cm}
The performance obtained by the studied systems on the development and evaluation data are shown in Table \ref{tab:development_results}. As expected, using the MSE error homogeneously consistently results in much better performance than the MSE+CE combinations. In addition, the proposed system outperforms the DCASE 2020 SELD baseline across the evaluation metrics, particularly on the SED metrics. This is most likely due to the underfitting effect on the SED subtask of the baseline, making it underperforming on this subtask. Overall, using FOA and MIC data, the proposed system reduces the combined SELD error by $0.08$ and $0.11$ absolute on the development data from that of the baseline, respectively. The corresponding error reduction by {\bf Submission 2} on the evaluation data reaches $0.11$ and $0.09$, respectively.

Our submission to the DCASE 2020 Task 3 was ranked 6$^{\text{th}}$ overall. This is an encouraging result given that the submission systems were compact and neither relied on ensemble nor multiple microphone arrays \footnote{\scriptsize http://dcase.community/challenge2020/task-sound-event-localization-and-detection-results}.

%

\vspace{-0.15cm}
\section{Conclusions}
\vspace{-0.15cm}
This work investigated the loss functions used for SELD multitask modeling. We showed empirical evidence that the combination of the sigmoid CE loss (for the SED subtask) and the MSE loss (for the DOA estimation subtask), which is commonly used, often results in underfitting effect on the former. As an alternative, when the two subtasks were formulated as regression problems and the MSE loss was used for both, the multitask network was able to converge better, resulting in better and balanced performance. Experimental results on the development and evaluation set of the DCASE 2020 SELD task showed significant improvements over the DCASE 2020 baseline across all the evaluation metrics.

\bibliographystyle{IEEEtran}
\bibliography{refs}

\end{sloppy}
\end{document}